\documentclass[twocolumn,floatfix,prb,aps,showpacs]{revtex4}
\usepackage{graphicx,amsmath,amssymb}
\usepackage{nicefrac}

\newcommand{\PRL}{Phys.\ Rev.\ Lett.}
\newcommand{\PRB}{Phys.\ Rev.\ B}

\newcommand{\be}{\begin{equation}}
\newcommand{\ee}{\end{equation}}
\newcommand{\bnu}{\bar{\nu}}

\begin{document}

\title{Phase Diagram of the Two-Component Fractional Quantum Hall Effect}
\author{Alexander C. Archer and Jainendra K. Jain}
\affiliation{Department of Physics, 104 Davey Lab, Pennsylvania State University, University Park PA, 16802, USA}

\date{\today}
\begin{abstract}
We calculate the phase diagram of the two-component fractional quantum Hall effect as a function of the spin or valley Zeeman energy and the filling factor, which reveals new phase transitions and phase boundaries spanning many fractional plateaus. This phase diagram is relevant to the fractional quantum Hall effect in graphene and in GaAs and AlAs quantum wells, when either the spin or the valley degree of freedom is active. 
\end{abstract}
\maketitle

The interplay between the Coulomb interaction and the electron spin degree of freedom has led to an impressive amount of physics in the fractional quantum Hall effect (FQHE). Phase transitions have been observed as a function of the Zeeman splitting, $E_{\rm Z}$, in transport,\cite{Eisenstein,Duspin,Yeh,Cho,Eom, Betthausen} optical \cite{Kukushkin,Yusa01,Gros07,Hayakawa} and NMR experiments.\cite{Melinte,Smet01,Kraus,Smet,Tiemann} 
Recent years have witnessed a remarkable resurgence of interest in multicomponent FQHE due to the experimental observation of FQHE in systems with both spin and valley degrees of freedom, such as AlAs quantum wells,\cite{Shayegan1,Shayegan2,Shayegan3} graphene,\cite{Kim,Yacoby,Yacoby2} and H-terminated Si(111) surface.\cite{Kane12} These enable new and more powerful methods of controlling the relative strengths of the (spin or valley) ``Zeeman" splitting and the interaction, thus opening the door into investigations of the physics of multicomponent FQHE states over a broad range of parameters. 

We consider FQHE for SU(2) electrons, applicable to parameter regimes in which either the valley or the spin degree of freedom is active. For simplicity, we will refer to the two-components generically as ``spins." Phase transitions at the isolated filling factors $\nu=n/(2pn\pm 1)$, $n$ and $p$ integers, were studied theoretically previously.\cite{Wu93,Park98,Park982,Park01,Dav12} We obtain in this Letter the more complete $E_{\rm Z}-\nu$ phase diagram, which reveals many phase boundaries arising from a competition between the Zeeman and the Coulomb energies.

\begin{figure}[b]
\includegraphics[scale=0.5]{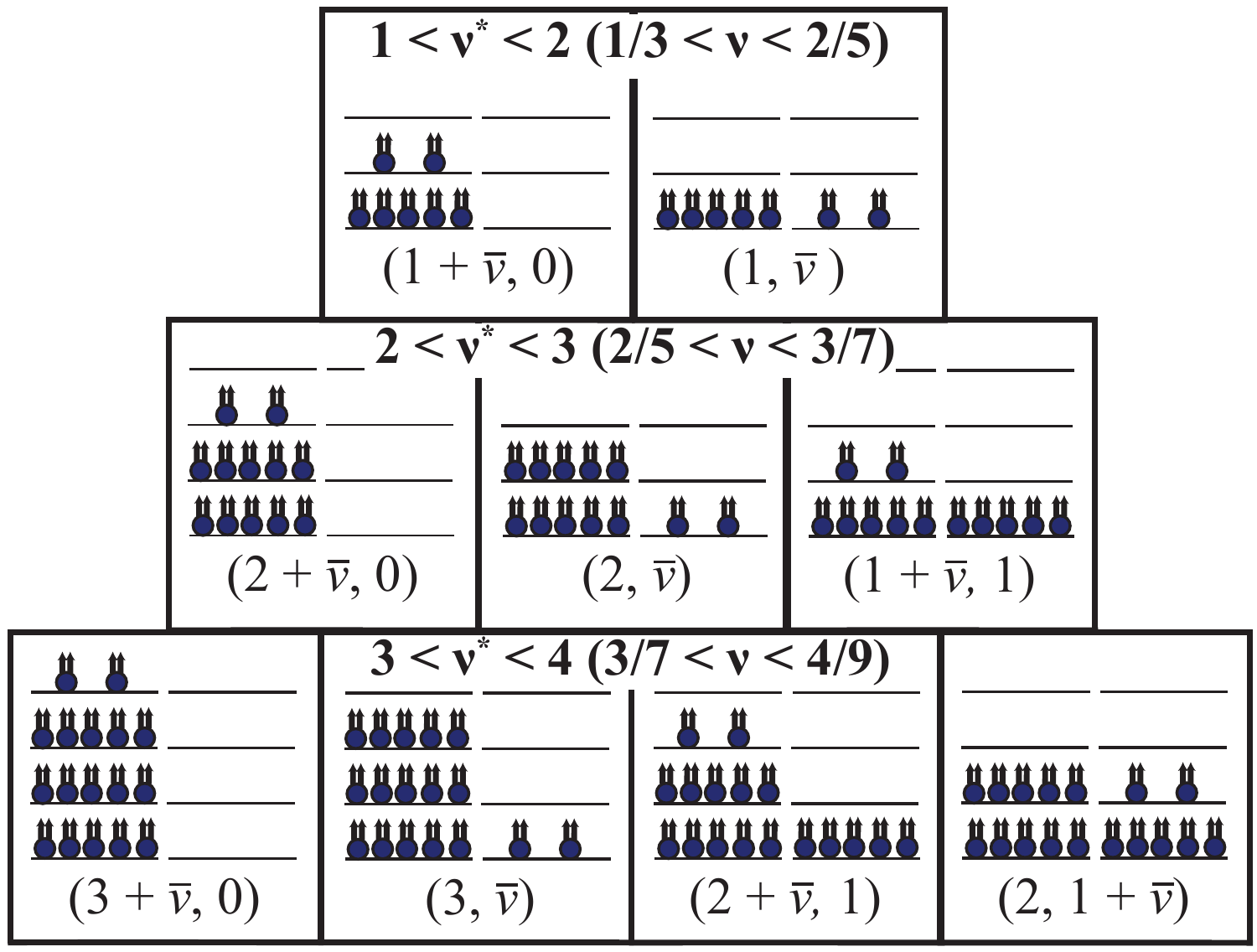}
\caption{Schematic depiction of the states with different polarizations in the filling factor range $1<\nu^*<2$ (top), $2<\nu^*<3$ (middle) and $3<\nu^*<4$ (bottom). Composite fermions are pictured as electrons carrying two arrows (representing bound vortices). The horizontal lines represent the CF $\Lambda$Ls, with the spin-up $\Lambda$Ls on the left and spin-down $\Lambda$Ls on the right. The states are labeled $(\nu_{\uparrow}^*,\nu_{\downarrow}^*)$; the filling factor of the partially filled $\Lambda$ level is $\bnu$.}
\label{schematic}
\end{figure}

In the filling factor range of interest below, electrons capture two vortices to transform into composite fermions (CFs).\cite{Jain89,Jainbook,HLR,Lopez} Composite fermions fill $\nu^*$ Landau-like levels called $\Lambda$ levels ($\Lambda$Ls), where $\nu$ and $\nu^*$ are related by $\nu=\nu^*/(2\nu^*\pm 1)$. In general, $\nu_{\uparrow}^*$ spin-up and $\nu_{\downarrow}^*$ spin-down $\Lambda$Ls are occupied, with $\nu^*=\nu_{\uparrow}^*+\nu_{\downarrow}^*$. The corresponding state, labeled $(\nu_{\uparrow}^*,\nu_{\downarrow}^*)$, has spin polarization $P=(\nu_{\uparrow}^*-\nu_{\downarrow}^*)/\nu^*$. (We take below the convention $\nu_\uparrow^*\geq \nu_\downarrow^*$.)  Which state is favored is an energetic question that requires a precise quantitative understanding of the various states. We begin by making certain simplifying assumptions.  We will assume that only one of the $\Lambda$Ls is partially occupied, so one of $\nu_{\uparrow}^*$ and $\nu_{\downarrow}^*$ is an integer and the other will be written as $j+\bnu$, where $0\leq \bnu \leq 1$ is the filling factor of the partially filled $\Lambda$L.  This assumption is valid for weakly interacting composite fermions. We neglect the Skyrmion physics\cite{Sondhi}; as discussed below, it is not relevant at the phase boundaries of interest in this work.  We also  neglect disorder and Landau level (LL) mixing. Fig.~\ref{schematic} schematically depicts all the states between 1/3 and 4/9 that we have studied below. 

To proceed further, we need a realistic model for the state of composite fermions in the partially filled $\Lambda$L, which is dictated by the weak residual interaction between composite fermions. For small $\bnu$ (small $1-\bnu$), we expect the formation of a crystal of CF particles (CF holes). We will model the entire range $0<\bnu<1$ as a crystal, and for completeness, we will consider both the particle and hole crystals for the entire range.The validity of various approximations is discussed below.

With these assumptions, it is possible to formulate a wave function for the FQHE state at $\nu$ following standard methods.\cite{Jain89}  We fist construct the wave function $\chi_{\nu^*}$ of {\em electrons} at $\nu^*$:
\be
\chi_{\nu^*}= A[\Phi_{\nu_\uparrow^*,\nu_\downarrow^*}\{z_1,\cdots  z_{N}  \}u_1 \cdots u_{N_\uparrow} d_{N_\uparrow+1} \cdots d_N]
\ee
\be
\Phi_{\nu_\uparrow^*,\nu_\downarrow^*}=\Phi_{\nu_\uparrow^*}\{z_1,\cdots z_{N_\uparrow}  \}\Phi_{\nu_\downarrow^*} \{z_{N_\uparrow+1}\cdots z_{N}  \}
\label{IQHE}
\ee
Here $z_j=x_j+iy_j$ denotes the position of an electron; $u$ and $d$ are the up and down spin spinors; $A$ indicates antisymmetrization; and $\Phi_{\nu_\uparrow^*}$ and $\Phi_{\nu_\downarrow^*}$ are wave functions of spin up and spin down electrons at $\nu_\uparrow^*$ and $\nu_\downarrow^*$.
One of the two factors on the right hand side of Eq.~\ref{IQHE} corresponds to integer filling, and is thus a single Slater determinant. We have two possible choices for the other factor at filling $j+\bnu$: (i) crystal of electrons on top of $j$ filled LLs, or (ii) crystal of holes on top of $j+1$ filled LLs. Because the Coulomb interaction commutes with spin, the many-body eigenstates must be eigenstates of $\vec{S}^2$. For the states relevant to our current study (Fig. \ref{schematic}), all occupied orbitals of spin-down electrons in $\Phi_{\nu_\downarrow^*}$ are definitely occupied for spin-up electrons $\Phi_{\nu_\uparrow^*}$; the product wave function $\Phi_{\nu_\uparrow^*,\nu_\downarrow^*}$ is therefore annihilated by the spin raising operator. In other words, it satisfies the Fock condition,\cite{Fock} which guarantees that it is an eigenstate of both $S_z$ and $\vec{S}^2$ with $S=S_z$.

\begin{figure}[t]
\includegraphics[scale=.6]{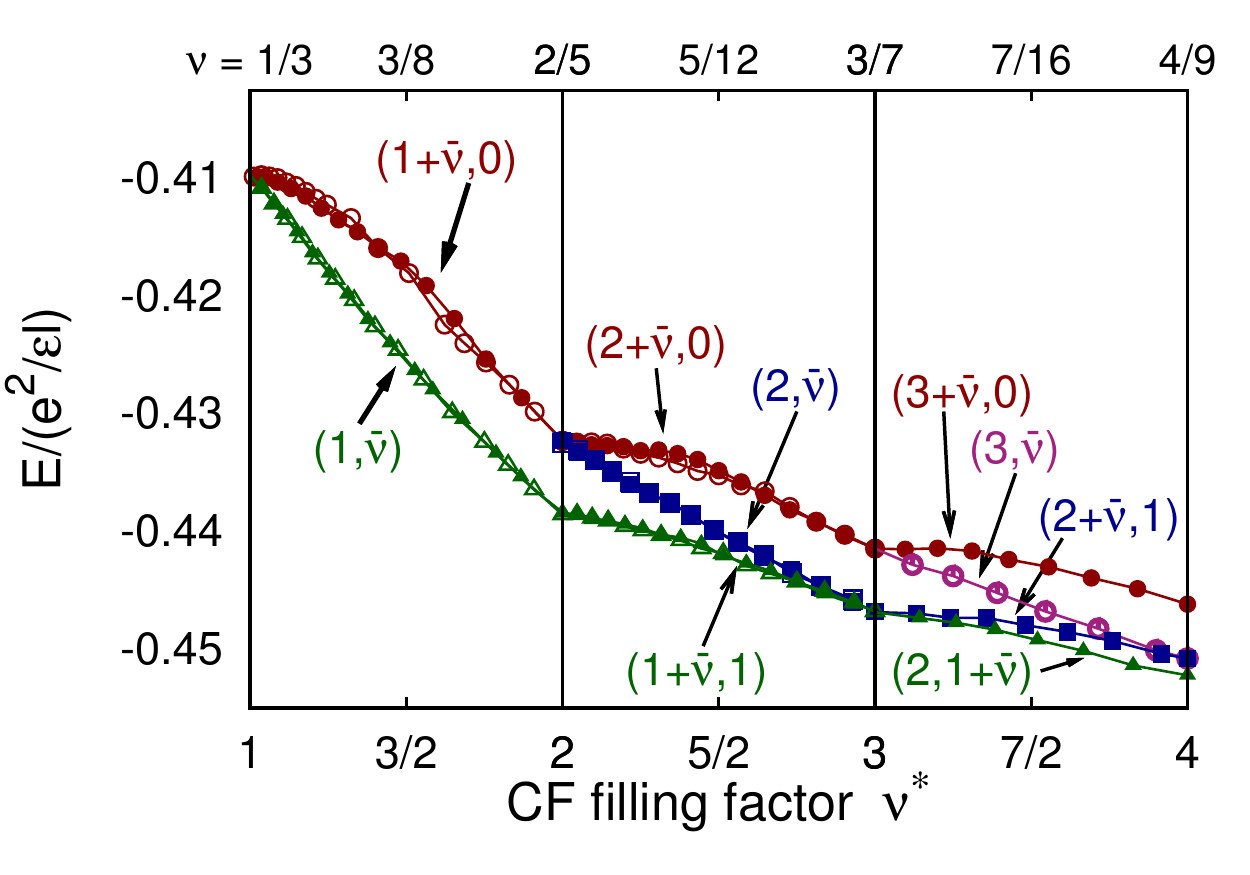}
\caption{The energy per particle for various states $(\nu_{\uparrow}^*,\nu_{\downarrow}^*)$ in the filling factor range $1<\nu^*<4$. The energy $E$ is quoted in units of $e^2/\epsilon \ell$, where $\ell=\sqrt{\hbar c/eB}$ is the magnetic length and $\epsilon$ is the dielectric constant of the host material, and includes interaction with a uniform neutralizing background. The dark (open) symbols correspond to states in which the partially filled $\Lambda$L contains a crystal of CF particles (holes). 
}
\label{Etot}
\end{figure}

We next composite-fermionize this wave function to construct the trial wave functions for the FQHE state at $\nu$:
\be
\Psi_{\nu}= A[\Psi^{\rm CF}_{\nu_\uparrow^*,\nu_\downarrow^*}\{z_1,\cdots  z_{N}  \}u_1 \cdots u_{N_\uparrow} d_{N_\uparrow+1} \cdots d_N]
\ee
\be
\Psi^{\rm CF}_{\nu_\uparrow^*,\nu_\downarrow^*}=P_{\rm LLL} \prod_{j<k=1}^N(z_j-z_k)^2 \Phi_{\nu_\uparrow^*,\nu_\downarrow^*}
\ee
Here the Jastrow factor $\prod_{j<k=1}^N(z_j-z_k)^2$ attaches two vortices to each electron to convert it into a composite fermion, and $P_{\rm LLL}$ denotes lowest Landau level (LLL) projection, evaluated using the method described in Ref.~\onlinecite{JK}. The spin quantum numbers are preserved\cite{Jainbook} under composite fermionization, guaranteeing that the wave function $\Psi_{\nu}$ also has proper symmetry under rotation in spin space. The determination of phase diagram requires an evaluation of the interaction energy
\be
E={\langle \Psi_{\nu}| V |  \Psi_{\nu} \rangle \over \langle \Psi_{\nu}|  \Psi_{\nu} \rangle}={\langle \Psi^{\rm CF}_{\nu_\uparrow^*,\nu_\downarrow^*}| V | \Psi^{\rm CF}_{\nu_\uparrow^*,\nu_\downarrow^*} \rangle \over \langle \Psi^{\rm CF}_{\nu_\uparrow^*,\nu_\downarrow^*}|  \Psi^{\rm CF}_{\nu_\uparrow^*,\nu_\downarrow^*} \rangle}
\ee
where $V$ is the interaction, which will be evaluated by the Metropolis Monte Carlo method.

We will use the spherical geometry \cite{Haldane83} for our calculation, in which $N$ electrons are confined to the surface of a sphere with $2Q$ flux quanta passing through it. This geometry is convenient as it has no complications due to edges and enables a simple treatment of the interaction with the background charge (taken as a single positive charge at the center). A disadvantage appears, at first, to be that a triangular crystal cannot be wrapped around a sphere without creating defects. We overcome this by exploiting J.J. Thomson's famous plum pudding model of the atom,\cite{Thomson} wherein the locations of classical charged particles on a sphere are obtained by minimizing their Coulomb energy. (These positions have been obtained by powerful numerical techniques, and tabulated in the literature.\cite{thommin}) As one may expect, this is essentially a triangular crystal with a few defects introduced by the curvature of the spherical surface. In the thermodynamic limit one expects the effect of the defects to be insignificant. We study a system with $N=84$ particles for our calculations below, varying the filling factor by considering different possible values of $2Q$. Our results represent the thermodynamic limit, as seen by noting that the critical Zeeman energies for 2/5, 3/7 and 4/9 agree with those obtained previously by explicitly evaluating the thermodynamic limit.\cite{Park98,Park01}  Technical details of the construction of the CF particle and CF hole crystals in the spherical geometry are given in the Supplemental Material.\cite{SM}

\begin{figure}
\includegraphics[scale=.6]{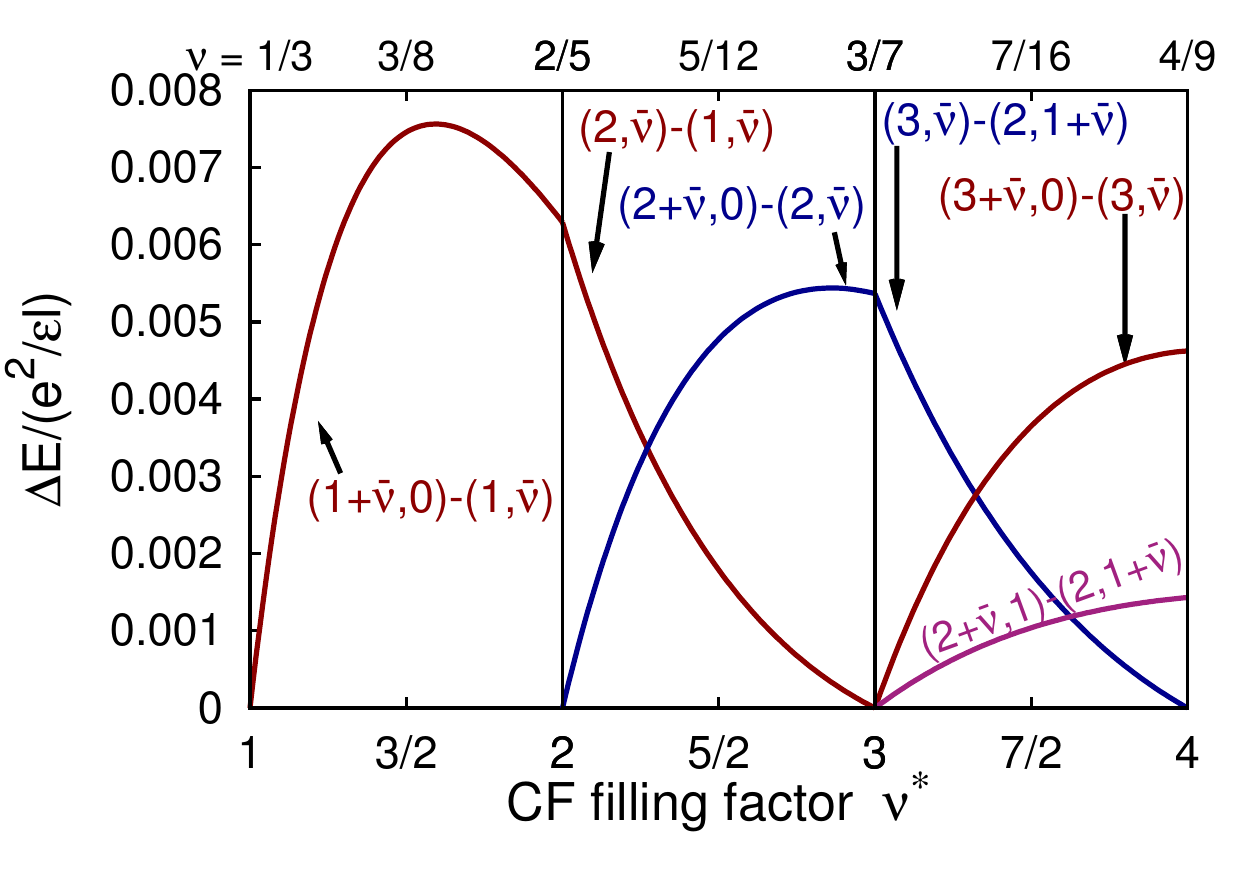}
\caption{Difference between the energies (per particle) of the states with different polarizations, shown near each line, in the filling factor range $1<\nu^*<4$.}
\label{Edif}
\end{figure}

\begin{figure}
\includegraphics[scale=.7]{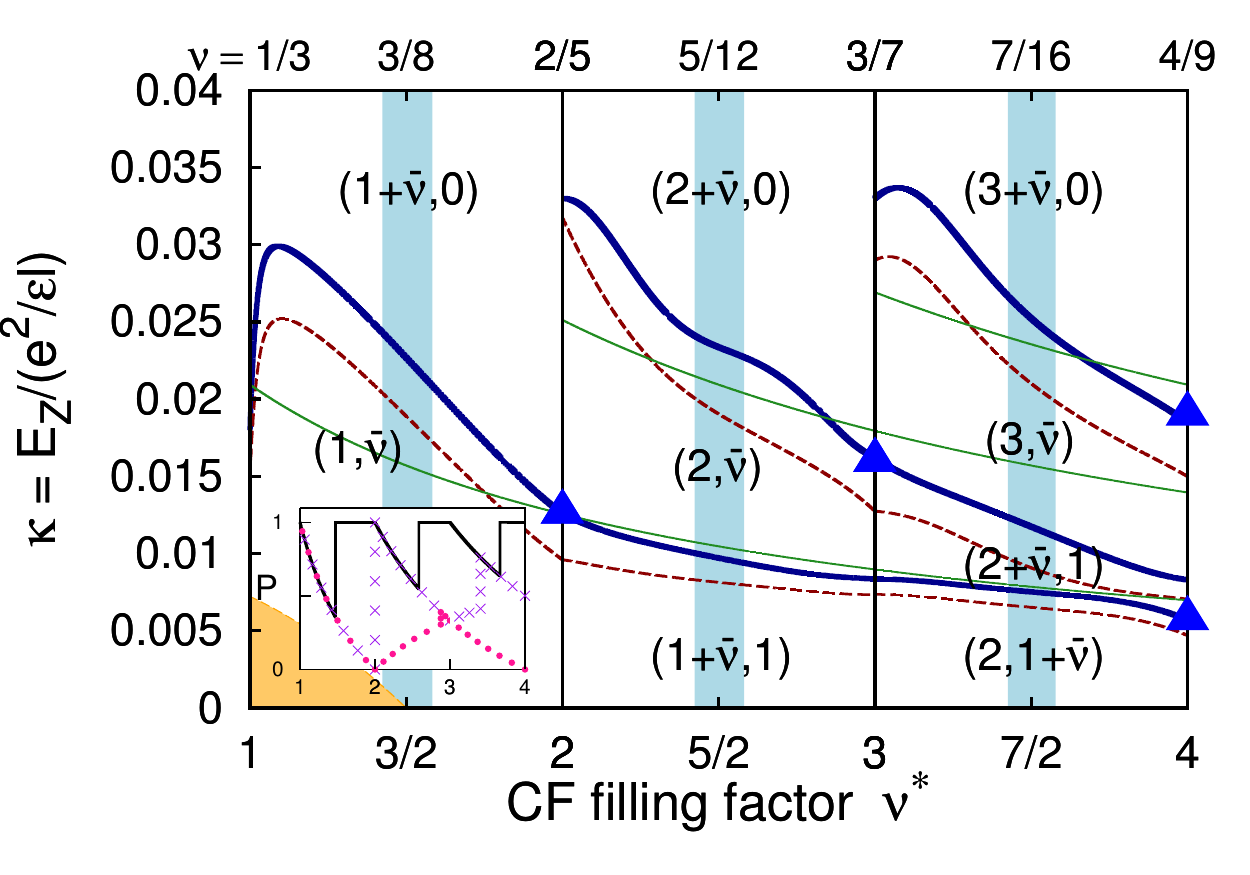}
\caption{The calculated polarization phase diagram of various states $(\nu_{\uparrow}^*,\nu_{\downarrow}^*)$ in the CF filling factor range $1<\nu^*<4$ ($1/3<\nu<4/9$) as a function of $\kappa=E_{\rm Z}/(e^2/\epsilon \ell)$. The thick blue lines are for a system of zero thickness; the dashed red lines correspond to a quantum well width of 40 nm for density $10^{11}\text{cm}^{-2}$; and thin green lines indicate the phase boundaries predicated by a model of noninteracting composite fermions. The blue triangles indicate the spin phase transitions at $\nu^*=n$ identified previously.\cite{Park98} The yellow shade depicts the region where Skyrmions are present, and the vertical blue shaded regions depict fillings where transitions take place from one fraction to the next. The inset shows the polarization $P=(\nu_{\uparrow}^*-\nu_{\downarrow}^*)/\nu^*$ as a function of $\nu^*$ for fixed values of $\kappa=0.0225$ (black solid line), 0.0125 (purple crosses) and 0.005 (red circles), as evaluated for a zero thickness sample.}
\label{Ecrit}
\end{figure}

Fig. \ref{Etot} shows the Coulomb energies of the various states as a function of the filling factor in the range $1/3<\nu<4/9$. The filled (open) symbols represent the energies for states where the partially filled $\Lambda$L is modeled as a crystal of CF particles (holes). (In the range $3/7 < \nu < 4/9$ only the CF particle crystal has been evaluated.) The energies include the interaction with the uniform positive background. The energy difference between the CF particle and hole crystals for a given polarization $P$ is small compared to that between states with different polarizations.  Fig. \ref{Edif} depicts the energy differences between two consecutive states, where we take the lower of the particle or hole crystal energies for a given spin polarization. From the difference, it is straightforward to determine the critical $E_{\rm Z}\equiv \kappa (e^2/\epsilon \ell)$ for the various transitions, plotted in Fig. \ref{Ecrit}. (In both Figs. \ref{Edif} and \ref{Ecrit} the curves have been smoothened to eliminate fluctuations due to finite system size. The uncertainty in the curves due to finite size effects is estimated to be $\delta\kappa\approx 0.001$.) The phase diagram in Fig. \ref{Ecrit} is the main result of our Letter.  

A striking feature of the phase diagram is the presence of lines that extend across many FQHE plateaus. 
For sufficiently large $\kappa$, the state is fully polarized at all fillings, as expected.  Interestingly, for a range of $\kappa$ values, the state at two consecutive fractions along the sequences $n/(2n+1)$ is fully spin polarized, but the state at intermediate fillings is partially polarized. The measurements of Tiemann {\em et al.} clearly show such behavior where both 1/3 and 2/5 are fully spin polarized but the intermediate state is partially polarized. The inset in Fig. \ref{Ecrit} shows the spin polarization $P=(\nu_{\uparrow}^*-\nu_{\downarrow}^*)/(\nu_{\uparrow}^*+\nu_{\downarrow}^*)$ as a function of $\nu$ for several fixed values $\kappa$ (as would be the case when $\nu$ is varied by changing the density at a fixed $B$), indicating complex $\kappa$ dependent behavior.  The blue triangles were obtained previously in Ref.~\onlinecite{Park98} and exhaust all spin polarization transitions at $\nu^*=$ integer. These are now seen to continue, as $\nu^*$ is varied, along phase boundaries that go ``sideways" (as opposed to connecting the adjacent blue triangles). Additional phase transitions also appear as $\nu^*$ is varied away from integer values of $\nu^*$, although they involve only a small change in spin polarization close to $\nu^*=$ integer.

A free-CF model provides useful insight into the structure of the phase diagram. This model considers noninteracting composite fermions, with the CF cyclotron energy taken as\cite{Duspin,Kukushkin,Melinte,Park98,HLR,Jainbook} $\hbar\omega_c^*=\frac{m_e}{m_p^*}\frac{\hbar\omega_c}{2\nu^*\pm1}\equiv {1\over \alpha (2\nu^*\pm1)}{e^2\over \epsilon \ell}$ at $\nu={\nu^*\over 2\nu^*\pm 1}$, where $\omega_c=eB/m_ec$ is the cyclotron energy, $m_e$ is the electron mass in vacuum, and $m_p^*$ is the CF ``polarization mass."\cite{Park98,Kukushkin,Melinte} (For GaAs parameters, we have $m_p^*/m_e=0.026 \alpha \sqrt{B[T]}$.) Phase transitions occur due to a competition between the CF cyclotron energy and the Zeeman energy. From an elementary calculation, this model predicts that the phase boundaries between the different polarizations are given by $E_{\rm Z}={\rm integer}\times \hbar\omega_c^*$:
\begin{equation}
\kappa={{\rm integer}\over \alpha (2\nu^*\pm1)}\;\;\;{\rm (free}\,{\rm CF}\,{\rm model)}
\end{equation}
where the integer goes from 1 to int($\nu^*$). 
We show the phase diagram predicted by the free-CF model in Fig.~\ref{Ecrit} (thin lines), fixing $\alpha=15.9$ by requiring coincidence at the blue triangle at $\nu=2/5$.  (The polarization mass is known to be much larger than the ``activation mass" relevant for transport experiments.\cite{Park98,Kukushkin,Melinte})
The free-CF model reproduces certain qualitative features of the phase diagram obtained from the microscopic theory. We stress that the free-CF model with an effective mass parameter should only be treated as providing an approximate intuitive interpretation of the accurate results obtained from the microscopic calculation; the free-CF model can sometimes fail qualitatively.\cite{Park982}

We have also studied the effect of finite thickness assuming a cosine shaped wave function in a square quantum well, which leads to a width dependent effective interaction.\cite{SM}  As an illustration, the phase diagram is also shown (red dashed lines) for a quantum well of width $40$ nm at a density of $10^{11}$ cm$^{-2}$. Lower densities and shorter widths produce a smaller deviation from the zero width results.

We now discuss the validity of the assumptions made in our calculation. (i) At $\kappa=0$, the excitations of 1/3 are CF analogs of the $\nu=1$ Skyrmions\cite{Sondhi} and involve a macroscopic number of spin flips.\cite{Kamilla} The Skyrmion size rapidly shrinks with $\kappa$, and quantitative estimates show\cite{Kamilla} that Skyrmions at 1/3 occur only for $\kappa<0.007$. Given that there are no Skyrmion excitations for the spin singlet 2/5, the Skyrmion region is roughly estimated to be that shaded in yellow in Fig.~\ref{Ecrit}. Here the phase will be a crystal of Skyrmions,\cite{Skcrystal} with a spin polarization less than $(1-\bnu)/(1+\bnu)$ by an amount that depends sensitively on $\kappa$. The phase boundaries calculated above are not affected by this physics, because they occur at relatively high values of $\kappa$ where Skyrmions are not relevant. (ii) One of our main assumptions has been to model the state in the partially filled $\Lambda$L as a crystal. This should be accurate for filling factors close to $\nu^*$=integer. How about other phases of composite fermions, such as their stripes or FQHE? For $\nu^*>2$, a FQHE of composite fermions is unlikely but a stripe phase is competitive;\cite{LeeScarolaJain2002} near $\bnu=1/2$, the energy per particle of the stripe phase is estimated\cite{LeeScarolaJain2002} to be below the crystal by $\sim$ 0.001 $e^2/\epsilon \ell$ or less, which is much smaller than the energy difference between states with different polarizations, and will therefore not affect the phase boundary obtained above significantly. The most interesting region is $1<\nu^*<2$ ($1/3<\nu<2/5$), where certain FQHE states such as 4/11, 5/13, and perhaps 3/8, are known to occur\cite{Pan03} in very pure samples. Inclusion of these states will distort the phase boundary near these fillings in interesting ways, but our theoretical understanding of these states\cite{other} is currently not at a level where quantitative statements can be made. 
(iii) For graphene and other multivalley systems, it will be important to consider both the spin and valley indices to bring out the full physics.\cite{Kim,Toke} The present work is applicable when one of those two degrees of freedom is frozen. (iv) Disorder, not included above, will affect our results in several ways. Most importantly, the first order phase transitions at the phase boundaries will turn into continuous percolation transitions in the presence of disorder, producing extended states that allow identification of the phase transition in transport experiments (below). Disorder will also affect the energies of the various states differently, and thereby modify the phase boundaries. Finally, disorder will create spatial variations in the filling factor, which will provide a correction to the ``ideal" polarization. For example, while the ideal 1/3 state is fully spin polarized for all $\kappa$, disorder will slightly diminish $P$ for $\kappa<0.03$. (v) LL mixing should be small at large magnetic fields, but at relatively low fields it will also influence the phase boundaries, because it will affect dissimilarly polarized states differently. 
LL mixing will affect the filling factor regions $0<\nu<1$ and $1<\nu<2$ differently. In the absence of LL mixing the latter region maps into $0<\nu<1$ holes in the LLL, which have the same interaction as electrons, and therefore the above physics applies exactly (with composite fermions being bound states of {\em holes} and vortices). In the presence of LL mixing, the renormalization of the interaction by LL mixing will be different for electrons and holes, and the quantitative differences between the spin phase diagrams in the two regions should serve as a useful test of our understanding of the role of LL mixing. A reliable treatment of LL mixing, however, is outside the scope of the present work.

In light of the preceding paragraph, the phase diagram in Fig.~\ref{Ecrit} is to be viewed as a first step. Further work will be required for a more precise determination of the phase boundaries. In particular, qualitative deviations from the phase diagram in Fig.~\ref{Ecrit} will be indicative of the formation of new correlated phases.

Transport measurements can identify the phase boundary through an $R_{xx}$ peak as a function of the $E_{\rm Z}$, which appears because of the presence of extended states (due to disorder) at the transition point. These $R_{xx}$ peaks are flanked by two states with the same quantized Hall resistance $R_{xy}$. (These should be distinguished from the peaks at \cite{Goldman} $\nu^*=n+1/2$, shown schematically as vertical shaded regions in Fig. ~\ref{Ecrit}, which indicate transitions between states with different values of the Hall resistance.) Such peaks have been seen in a number of experiments, both at the special fractions $\nu=n/(2n\pm 1)$ and slightly away from it.\cite{Duspin,Cho,Eom,Kraus} Krauss {\em et al.} \cite{Kraus} have detected a sharp $R_{xx}$ peak inside the resistance minima near 2/3 and 3/5, and determined a phase boundary of the type calculated here, clearly delineating states with various spin polarizations as a function of the filling factor. Unfortunately, a technical difficulty in dealing with reverse flux attachment makes the calculation of the phase diagram along the sequence $n/(2n-1)$ much more challenging,\cite{Wu93,Dav12} and outside the scope of the present work. However, the phase transition at 2/3 is found to occur in Ref.~\onlinecite{Kraus} at $\sim 10$T; according to the free-CF model, this corresponds to $\alpha\sim 19$, which is in the same ballpark as the value obtained above (and, in fact, somewhat higher, as expected from finite width corrections). Resistance spikes have also been observed at magnetic phase transitions in the integer quantum Hall regime.\cite{Poortere} We also note that because the phase boundary involves a change of magnetization, it should exhibit hysteretic behavior, as observed previously.\cite{Cho,Eom,Smet01,Poortere}

We thank  B. E. Feldman, Y. Liu,  M. Shayegan, and A. Yacoby for many insightful discussions. We thank the National Science Foundation for support under grant no. DMR-1005536, and the Research Computing and Cyberinfrastructure, a unit of Information Technology Services at Pennsylvania State University, for providing high-performance computing resources.


\begin{thebibliography}{0}



\bibitem{Eisenstein} J. P. Eisenstein, H. L. Stormer, L. Pfeiffer, and K. W. West, Phys. Rev. Lett. {\bf 62}, 1540 (1989).

\bibitem{Duspin} R. R. Du {\em et al.}, Phys. Rev. Lett. {\bf 75}, 3926 (1995).

\bibitem{Cho} H. Cho {\em et al.}, \PRL\ \textbf{81}, 2522 (1998).

\bibitem{Yeh} A. S. Yeh {\em et al.}, Phys. Rev. Lett. {\bf 82}, 592 (1999)

\bibitem{Eom} J. Eom {\em et al.}, Science {\bf 289}, 2320 (2000).

\bibitem{Betthausen} B. A. Piot {\em et al.}, \PRB\ {\bf 82}, 081307 (2010); 
C. Betthausen {\em et al.}, preprint.

\bibitem{Kukushkin}
I. V. Kukushkin, K. v. Klitzing and K. Eberl, Phys. Rev. Lett. {\bf 82}, 3665 (1999); I. V. Kukushkin, J. H. Smet, K. von Klitzing, and K. Eberl, {\em ibid.} {\bf 85}, 3688 (2000).

\bibitem{Yusa01} G. Yusa, H. Shtrikman, and I. Bar-Joseph, Phys. Rev. Lett. {\bf 87}, 216402 (2001).

\bibitem{Gros07} G. Groshaus, P. Plochocka-Polack, M. Rappaport, V. Umansky, I. Bar-Joseph, B. S. Dennis, L. N. Pfeiffer, K. W. West, Y Gallais, and A. Pinczuk, Phys. Rev. Lett. {\bf 98}, 156803 (2007).

\bibitem{Hayakawa} J. Hayakawa, K. Muraki and G. Yusa, Nat. Nano. {\bf 8}, 31 (2013).

\bibitem{Melinte}
S. Melinte {\em et al.}, Phys. Rev. Lett.  {\bf 84}, 354 (2000);
N. Freytag {\em et al.}, {\em ibid.} {\bf 89}, 246804 (2002).

\bibitem{Smet01} J. H. Smet {\em et al.}, Phys. Rev. Lett. {\bf 86}, 2412 (2001).

\bibitem{Kraus} S. Kraus {\em et al.}, Phys. Rev. Lett. {\bf 89}, 266801 (2002).

\bibitem{Smet} J. H. Smet {\em et al.}, Phys. Rev. Lett. {\bf 92}, 086802 (2004).

\bibitem{Tiemann} L. Tiemann, G. Gamez, N. Kumuda and K. Muraki, Science {\bf 335}, 828 (2012).

\bibitem{Shayegan1} N. C. Bishop, M. Padmanabhan, K. Vakili, Y.P. Snokolnikov, E. P. De Poortere, and M. Shayegan, Phys. Rev. Lett. {\bf 98}, 266404 (2007).

\bibitem{Shayegan2} M. Padmanabhan, T. Gokmen, M. Shayegan, Phys. Rev. B {\bf 80}, 035423 (2009); Phys. Rev. B {\bf 81}, 113301 (2010).

\bibitem{Shayegan3} T. Gokmen, M. Padmanabhan, and M. Shayegan, Nature Physics {\bf 6}, 621 (2010).

\bibitem{Kim} C. R. Dean {\em et al.}, Nat. Phys. {\bf 7}, 693 (2011).

\bibitem{Yacoby} B. E. Feldman, B. Krauss, J. H. Smet and A. Yacoby, Science {\bf 337}, 1196 (2012).

\bibitem{Yacoby2} Benjamin E. Feldman, Andrei J. Levin, Benjamin Krauss, Dmitry Abanin, Bertrand. I. Halperin, Jurgen H. Smet, Amir Yacoby, cond-mat arXiv:1303.0838 (2013).

\bibitem{Kane12} T.M. Kott, B. Hu, S. H. Brown, and B. E. Kane, arXiv:1210.2386 (2012).

\bibitem{Wu93} X.G. Wu, G. Dev, and J.K. Jain, Phys. Rev. Lett. {\bf 71}, 153 (1993).

\bibitem{Park98} K. Park and J. K. Jain, Phys. Rev. Lett. {\bf 80}, 4237 (1998).

\bibitem{Park982} K. Park and J. K. Jain, Phys. Rev. Lett.  {\bf 83}, 5543 (1999).

\bibitem{Park01} K. Park and J. K. Jain,  Solid State Commun. {\bf 119}, 291 (2001).

\bibitem{Dav12} S.C. Davenport and S.H. Simon, Phys. Rev. B {\bf 85}, 245303 (2012).

\bibitem{Jain89}  J.K. Jain, Phys. Rev. Lett. {\bf 63}, 199 (1989);
Phys. Rev. B {\bf 41}, 7653 (1990).

\bibitem{Jainbook} Jainendra K. Jain, {\it Composite Fermions} (Cambridge University Press, Cambridge, UK, 2007).

\bibitem{HLR}  B.~I.~Halperin, P.~A.~Lee, N.~Read, \PRB\ \textbf{47} 7312 (1993).

\bibitem{Lopez} A. Lopez and E. Fradkin, Phys. Rev. B {\bf 44}, 5246 (1991).


\bibitem{Sondhi} S.L. Sondhi, A. Karlhede, S. A. Kivelson, and E. H. Rezayi, Phys. Rev. B {\bf 47}, 16419 (1993); H.A. Fertig, L. Brey, R. C\^ote, and A. H. MacDonald, Phys. Rev. B {\bf 50}, 11018 (1994).

\bibitem{JK} J.K. Jain and R.K. Kamilla, Phys. Rev. B {\bf 55},
R4895 (1996); Int. J. Mod. Phys. B {\bf 11}, 2621 (1997).

\bibitem{Fock} M. Hamermesh, {\em Group Theory and Its Applications to Physical Problems} (Dover Publications, 1989).

\bibitem{Haldane83} F.D.M. Haldane, Phys. Rev. Lett. {\bf 51}, 605 (1983).

\bibitem{Thomson} J.J. Thomson, Philos. Mag. {\bf 7}, 237 (1904).

\bibitem{thommin} David J. Wales, Hayley McKay, and Eric L. Altschuler, Phys. Rev. B {\bf 79}, 224115 (2009); David J. Wales and Sidika Ulker, Phys. Rev. B {\bf 74}, 212101 (2006); M. J. Bowick, C. Cecka, L. Giomi, A. Middleton and K. Zielnicki, http://thomson.phy.syr.edu/thomsonapplet.php.

\bibitem{SM} See Supplemental Material for technical details of the wave function construction in the spherical geometry.

\bibitem{Kamilla} R. K. Kamilla, X. G. Wu and J. K. Jain, Solid State Commun. {\bf 99}, 289 (1996); R. L. Doretto {\em et al.}, Phys. Rev. B {\bf 72}, 035341 (2005).

\bibitem{Skcrystal} L. Brey, H. A. Fertig, R. C\^ote, and A. H. MacDonald, Phys. Rev. Lett. {\bf 75}, 2562 (1995).

\bibitem{LeeScarolaJain2002} S.~Y.~Lee,V.~W.~Scarola, and J.~K.~Jain,  \PRL\ \textbf{87}, 256803 (2001); \PRB\ \textbf{66}, 085336 (2002).

\bibitem{Pan03} W. Pan, H. L. Stormer, D. C. Tsui, L. N. Pfeiffer, K. W. Baldwin, and K. W. West, Phys. Rev. Lett. {\bf 90}, 016801 (2003).

\bibitem{other} V.~W.~Scarola, J.~K.~Jain, E.~H.~Rezayi, \PRL\, \textbf{88}, 216804 (2002); C.-C. Chang, S. S. Mandal and J. K. Jain, \PRB\ \textbf{67}, 121305(R) (2003); M. O. Goerbig, P.Lederer, and C. Morais Smith, Phys. Rev. B {\bf 69}, 155324 (2004); A. W\'ojs, K.-S. Yi, and J. J. Quinn, \PRB\ \textbf{69}, 205322 (2004); A. W\'ojs, G. Simion, and J. J. Quinn, \PRB\ \textbf{75}, 155318 (2007); S. Mukherjee {\em et al.}, \PRL\ {\bf 109}, 256801 (2012).

\bibitem{Toke} K. Yang, S. Das Sarma and A. H. MacDonald, Phys. Rev. B {\bf 74}, 075423 (2006);
C. T\"oke and J. K. Jain, Phys. Rev. B {\bf 75}, 245440 (2007); 
M. O. Goerbig and N. Regnault, Phys. Rev. B {\bf 75}, 241405 (2007);
N. Shibata and K. Nomura, J. Phys. Soc. Japan \textbf{78}, 104708 (2009).

\bibitem{Goldman} V. J. Goldman, J. K. Jain and M. Shayegan,  Phys. Rev. Lett. {\bf 65}, 907 (1990).

\bibitem{Poortere} E. P. De Poortere, E. Tutuc, S. J. Papadakis, and M. Shayegan, Science {\bf 290}, 1546 (2000).


\end{thebibliography}
\end{document}